\documentclass[journal=jacsat]{achemso}
\usepackage[version=3]{mhchem} 
\usepackage{siunitx}
\usepackage{soul}
\usepackage{multirow}
\usepackage{xr}
\usepackage{float}
\usepackage{geometry}
\usepackage{graphicx}
\usepackage{dcolumn}
\usepackage{bm}
\usepackage{xcolor}
\usepackage{ccaption}
\usepackage{float}

\title{Adaptive Resetting for Informed Search Strategies and the Design of Non-equilibrium Steady-States}

\author{Tommer D. Keidar}
\affiliation[TAU]
{School of Chemistry, Tel Aviv University, Tel Aviv 6997801, Israel.}
\alsoaffiliation[BioSoft]
{The Center for Physics and Chemistry of Living Systems, Tel Aviv University, Tel Aviv 6997801, Israel.}
\altaffiliation{T.D.K. and O.B. contributed equally to this work}
\author{Ofir Blumer}
\affiliation[TAU]
{School of Chemistry, Tel Aviv University, Tel Aviv 6997801, Israel.}
\alsoaffiliation[BioSoft]
{The Center for Physics and Chemistry of Living Systems, Tel Aviv University, Tel Aviv 6997801, Israel.}
\altaffiliation{T.D.K. and O.B. contributed equally to this work}
\author{Barak Hirshberg}
\affiliation[TAU]
{School of Chemistry, Tel Aviv University, Tel Aviv 6997801, Israel.}
\email{hirshb@tauex.tau.ac.il}
\alsoaffiliation[BioSoft]
{The Center for Physics and Chemistry of Living Systems, Tel Aviv University, Tel Aviv 6997801, Israel.}
\alsoaffiliation[Sackler]
{The Center for Computational Molecular and Materials
Science, Tel Aviv University, Tel Aviv 6997801, Israel.}
\author{Shlomi Reuveni}
\email{shlomire@tauex.tau.ac.il}
\affiliation[TAU]
{School of Chemistry, Tel Aviv University, Tel Aviv 6997801, Israel.}
\alsoaffiliation[BioSoft]
{The Center for Physics and Chemistry of Living Systems, Tel Aviv University, Tel Aviv 6997801, Israel.}
\alsoaffiliation[Sackler]
{The Center for Computational Molecular and Materials
Science, Tel Aviv University, Tel Aviv 6997801, Israel.}

\begin{document}

\begin{abstract}
Stochastic resetting, the procedure of stopping and re-initializing random processes, has recently emerged as a powerful tool for accelerating processes ranging from queuing systems to molecular simulations. However, its usefulness is severely limited by assuming that the resetting protocol is completely decoupled from the state and age of the process that is being reset. We present a general formulation for state- and time-dependent resetting of stochastic processes, which we call adaptive resetting. This allows us to predict, using a single set of trajectories without resetting and via a simple reweighing procedure, all key observables of processes with adaptive resetting. These include the first-passage time distribution, the propagator, and the steady-state. Our formulation enables efficient exploration of informed search strategies and facilitates the prediction and design of complex non-equilibrium steady states, eliminating the need for extensive brute-force sampling across different resetting protocols. Finally, we develop a general machine learning framework to optimize the adaptive resetting protocol for an arbitrary task beyond the current state of the art. We use it to discover efficient protocols for accelerating molecular dynamics simulations.

\end{abstract}

\maketitle

\section*{Introduction}

Stochastic resetting has drawn significant scientific attention in recent years~\cite{Evans_majumdar_PRL,Evans_majumdar_JPhysA_review, gupta2022_review, kundu2024preface}  In it, a stochastic process is stopped at random times and restarted with independent and identically distributed initial conditions. The ability to create with it non-equilibrium steady states (NESS)~\cite{Evans_majumdar_PRL, Evans_majumdar_JPhysA_review, gupta2022_review, kundu2024preface, Eule_2016, Sarkar2022, Goerlich2024, Vatash2025}, and the relative ease at which stochastic resetting is implemented in lab conditions, made it a promising guinea pig for testing novel ideas in non-equilibrium statistical physics~\cite{Tal-Friedman2020, PRR_exp_reset, Faisant_2021_Exp, altshuler2023environmental, Sokolov2023}. Moreover, stochastic resetting can expedite first-passage processes~\cite{Evans_majumdar_JPhysA_review,Chechkin_Sokolov_Renewal, FPT_under_reset,inspection_paradox}, which has proven useful in accelerating computer algorithms, such as Las Vegas algorithms~\cite{Las_Vegas_Luby, las_Vegas_Alt}, search algorithms~\cite{resetting_search, Review_page_rank}, and molecular dynamics simulations~\cite{Ofir_JPCL, Ofir_NatureCom,non_exponential_kinetics}. Resetting is also central to our understanding of biological phenomena such as enzymatic catalysis and inhibition~\cite{Michaelis_Menten_PNAS, Robin2018}, and backtrack recovery by RNA polymerases~\cite{Backtrack_RNA, Particle_evap}.

The power and usefulness of the theory of stochastic resetting come from its ability to predict the time evolution, steady state, and first-passage time (FPT) distribution of a process undergoing restart, from its counterpart without resetting~\cite{Evans_majumdar_JPhysA_review, FPT_under_reset, Chechkin_Sokolov_Renewal, Eule_2016}. Despite its success,  a major shortcoming of the existing theory is that the resetting protocol is assumed to be independent of the state and progress of the underlying stochastic process. This severely limits the applicability of the theory, as we further explain below.  

\begin{figure}[t]
    \centering
    \includegraphics[width=0.9\linewidth
    ]{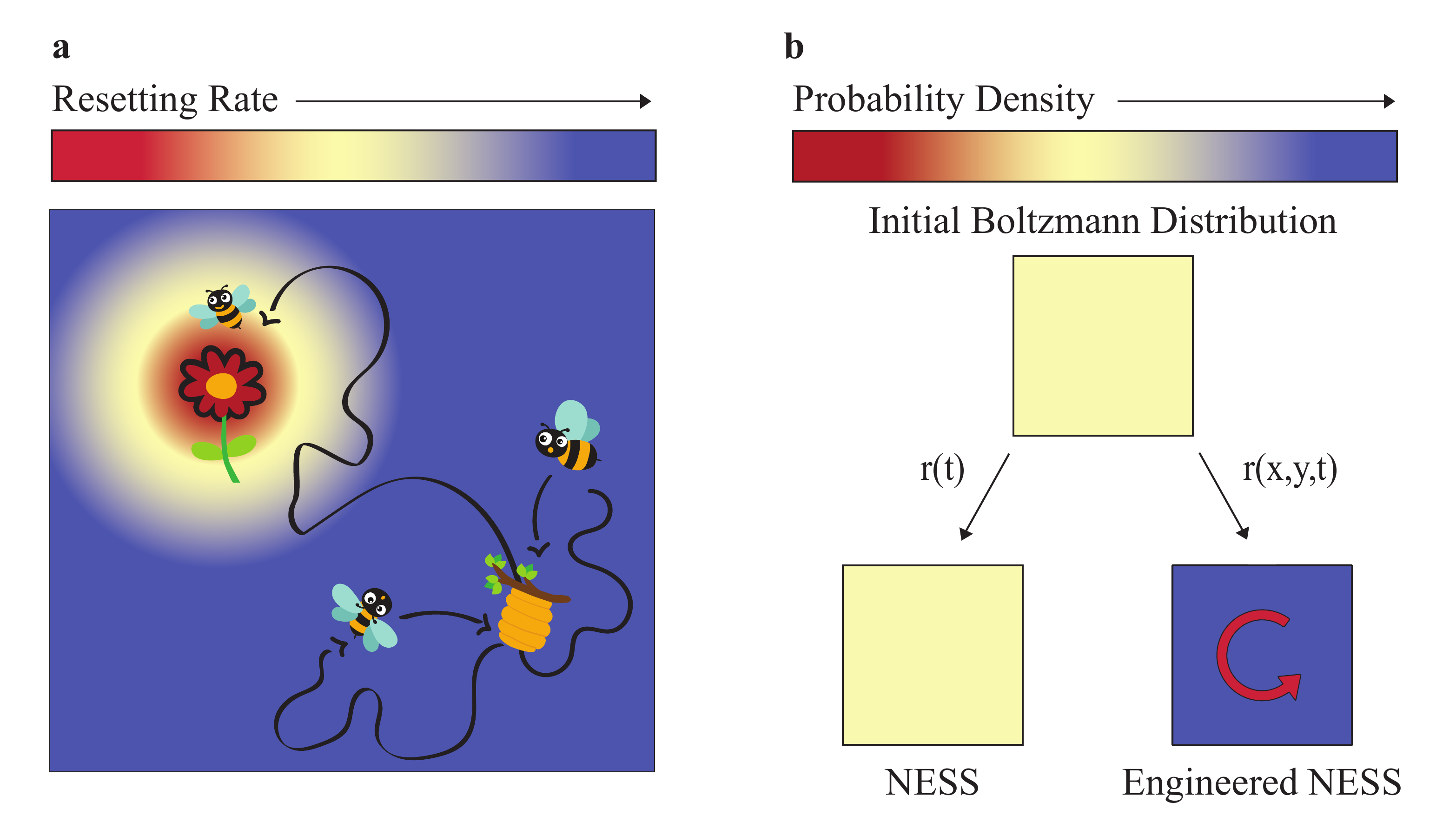}
    \caption{\textbf{a} Informed Search Strategies. To shorten search time, a foraging animal occasionally returns home (resetting). The animal adapts its resetting rate by gathering information on the proximity of food, e.g., by using the sense of smell. In this way, it avoids undesirable home returns, allowing it to locate food much faster. \textbf{b} NESS design via adaptive resetting. Starting from a uniform distribution (upper panel), resetting to the same initial conditions at a state-independent rate results in a uniform steady state (lower left panel). However, position-dependent resetting can produce non-uniform distributions, such as the curved arrow-shaped distribution in the lower right panel.}
    \label{fig: intro}
\end{figure}

Resetting is known to accelerate search processes under certain conditions \cite{Chechkin_Sokolov_Renewal,FPT_under_reset, Evans_majumdar_JPhysA_review,pal2020search}. However, since regular resetting is agnostic to the search \textit{progress}, it could also occur very close to the target. Preventing such undesirable resetting events is clearly beneficial. Searching agents may do this by sensing their proximity to the target, e.g., by smell, and adapt their resetting probability in response, as illustrated in Fig.~\ref{fig: intro}a. Similarly, when using resetting to expedite molecular simulations, adapting the resetting rate by including information on the reaction progress leads to substantial acceleration~\cite{Church2024}. 
All these examples fall outside the scope of the existing theory of stochastic resetting but can be addressed by accounting for a possible coupling between the state of the system and the resetting protocol. In these cases, the resetting rate adapts to the underlying dynamics in a state- and time-dependent manner, and we will thus refer to it as \textit{adaptive resetting}. 

Adaptive resetting also paves the way for using resetting as a tool for engineering NESS that could not be obtained with standard resetting. It is well known that resetting typically leads to a steady state \cite{Eule_2016, Evans_majumdar_JPhysA_review}. For example, resetting a particle diffusing in one dimension at a constant rate results in a Laplace distribution steady state \cite{Evans_majumdar_PRL}. However, a state-dependent resetting rate could lead to a broader range of possible outcomes. Imagine, for example, a particle diffusing in a two-dimensional box whose initial position is uniformly distributed. Stochastic resetting to this distribution at a state-independent rate will again result in a uniform, featureless, NESS. However, incorporating a position-dependent resetting rate can break symmetry and allow the design of more complicated NESS, as we illustrate in Fig. \ref{fig: intro}b.

So what is the key barrier hindering the development of a theory of adaptive, state- and time-dependent, resetting? For standard resetting, the dynamics of the probability distribution without resetting is sufficient to predict the behavior under restart \cite{Evans_majumdar_JPhysA_review}. For example, the probability distribution describing the position of a diffusing particle under resetting can be obtained directly from the fact that the probability distribution of a freely diffusing particle is Gaussian. As a result, the steady state under any distribution of resetting times can also be obtained. Similarly, the FPT distribution with resetting can be directly predicted from its counterpart without resetting \cite{FPT_under_reset}. Unfortunately,  this is no longer true for adaptive resetting, as the probability of reaching a certain point at a certain time without being reset depends on the entire history of the trajectory, via the state- and time-dependence of the resetting rate. 
In this case, one must account for all possible trajectory histories and their relative weights, which is a staggering challenge.

For overdamped Brownian motion under a conservative potential, Roldán and Gupta showed that solving a state-dependent resetting problem could be mapped to evaluating the quantum propagator of an auxiliary system with a potential that depends on the resetting protocol~\cite{PI_formalism_for_SR}. Essentially, they invoked the path integral formulation of quantum mechanics to sum over all possible trajectories. This can be done for a free particle or a quantum harmonic oscillator, where analytical solutions are known. Yet, analytical solutions for more complicated systems are notoriously hard to obtain, and numerical solutions scale exponentially with dimension. Other authors have thus focused on specific state-dependent resetting problems, giving \textit{ad hoc} solutions by circumventing calculation of the path integral~\cite{Evans_Majumdar_JPhysA_optimal_rate, Asymmetric_SR, Redner_reset_after_FP, Ali_2022, Energy_based_SR, Pinsky2020, Ye_2022, Singh2019, Berezhkovskii2017}.
This highlights the need in a general framework for stochastic processes with adaptive resetting, allowing to predict their behavior directly from trajectories without resetting, as is done for standard resetting.

Below, we present a general formulation of adaptive resetting. Practically, we show that a \textit{single} set of trajectories without resetting, obtained experimentally, numerically, or analytically, is sufficient to estimate all key observables with adaptive resetting via a simple reweighing procedure: the mean FPT (MFPT), the entire FPT distribution, the propagator, and the steady-state distribution.

We first demonstrate the power of our approach by analyzing informed search strategies, which utilize environmental information to lower the resetting rate in the vicinity of the target. There, we discover a cross-over in the optimal resetting strategy depending on the ratio of two length scales: the initial distance to the target and the range at which the target can be sensed effectively. Next, we use our approach to predict the tail of the NESS formed by a diffusing particle under a power-law resetting rate, which was so far out of reach theoretically. 
We also show how to use adaptive resetting to break the spatial symmetry of the underlying stochastic process and design complex and detail-rich NESS. 

Finally, we demonstrate how our framework can be used to optimize the adaptive resetting protocol to perform a given task. We do so by representing the state-dependent resetting probability as a neural network and training it to optimize a loss function that can be any observable of the process with adaptive resetting. The loss is calculated based on  trajectories without resetting using our reweighing procedure. As a concrete example, we find an optimized adaptive resetting protocol that minimizes the MFPT of conformational transitions in simulations of the mini-protein chignolin in explicit water.

\section{The MFPT with adaptive resetting}

We begin by defining a state- and time-dependent resetting rate $r(\boldsymbol{X}, t)$, with $\boldsymbol{X}$ being the state of the system and $t$ representing time. 
Given a trajectory, $\{ \boldsymbol{X}(t'), 0 \le t' \le t \}$, we can define a random variable describing its resetting time $R$ via its cumulative distribution function
\begin{equation}
\Pr(R\le t) = 1-\exp\left(-\int_{0}^tr(\boldsymbol{X}(t'), t')\,dt'\right). 
\label{probRlet}
\end{equation}
 Note that $R$ distributes differently for different trajectories under the \textit{same} functional choice of the resetting rate. 

We next construct the random variable describing the FPT under the resetting protocol $r(\boldsymbol{X}, t)$.
We observe that a first-passage process with resetting can be described in the following way: First, a trajectory is sampled, and this determines $T$, i.e., the FPT without resetting. The trajectory also sets the distribution of $R$ via equation (\ref{probRlet}).
We next sample $R$, and if $T\leq R$, we conclude that the FPT is simply $T$. If, however, $R < T$, resetting occurs before first-passage, a new trajectory is sampled, and the procedure is repeated, tallying $R$. Overall, the random variable describing the FPT under resetting is
\begin{equation}\label{ISR random variable}
    T_R=\begin{cases}
    T & \text{if $T \leq R$},\\
    R+T'_R & \text{if $R<T$},
    \end{cases}
\end{equation}
where $T'_R$ is an independent and identically distributed copy of $T_R$. Using the total expectation theorem, and averaging over all the trajectories, and all realizations of $R$, we find that the MFPT under restart is
\begin{equation}\label{ISR MFPT TET}
    \langle T_R\rangle=
    \left(\frac{1}{\Pr(T\leq R)}-1\right)\langle R|R<T\rangle
    + \langle T|T\leq R\rangle.
\end{equation}
From here, it is clear that the MFPT under restart can also be written as $\langle T_R\rangle=\frac{\langle min(T,R)\rangle}{\Pr(T\leq R)}$.  

\begin{figure}[t!]
     \centering
     \includegraphics[width=\linewidth]{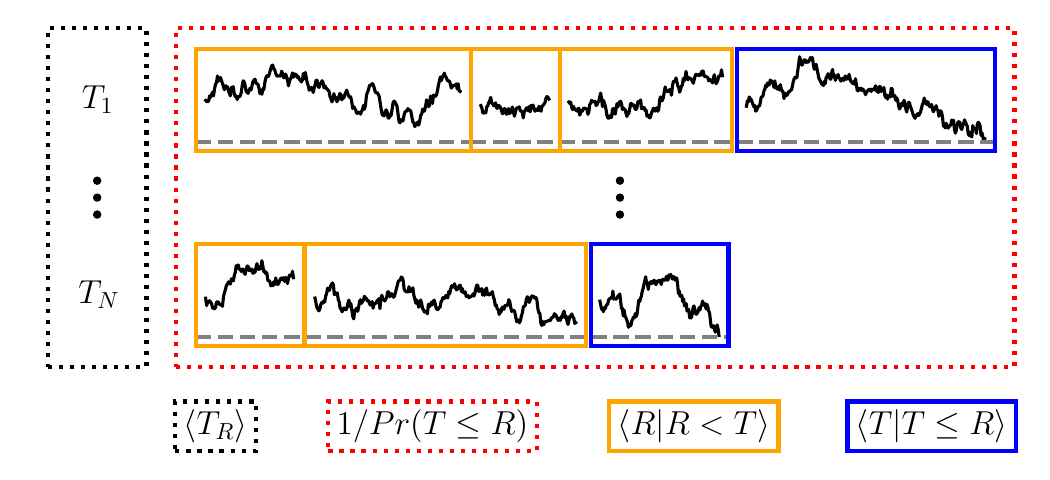}
 \caption{An illustration of trajectories with resetting. Each trajectory is composed of segments ending in resetting (yellow) and a final segment ending in first-passage to the gray dashed line (blue). The MFPT under restart $\langle T_R\rangle$ is given by equation (\ref{ISR MFPT TET}). It is the mean length of the blue segments $\langle T|T\leq R\rangle$, plus the mean length of the yellow segments $\langle R|R<T\rangle$, multiplied by the average number of yellow segments, $1/\Pr(T\leq R) -1$, in a single trajectory.}
     \label{fig:explainMFPT}
 \end{figure}

We can interpret equation (\ref{ISR MFPT TET}) as illustrated in Fig. \ref{fig:explainMFPT}. For a given trajectory with resetting, the first-passage is composed of several failed attempts to complete the process, each ending in resetting. These are followed by one successful attempt, ending in first-passage, and completing the process. The total number of attempts, failed plus successful, is geometrically distributed. Namely, all attempts are statistically independent with a success probability $\Pr(T\leq R)$. Therefore, the average number of attempts is $1/\Pr(T\leq R)$. The mean duration of an unsuccessful attempt is $\langle R|R<T\rangle$, and the mean duration of a successful one is $\langle T|T\leq R\rangle$. The MFPT with resetting is just the sum of the mean duration of a failed attempt times the mean number of failed attempts with the mean duration of the final successful attempt.

Equation (\ref{ISR MFPT TET}) is the first key result of this paper. It generalizes known results for the MFPT under resetting for a much broader class of resetting protocols. Usually, one assumes that the first passage and resetting processes are completely decoupled, making $T$ and $R$ statistically independent\cite{FPT_under_reset,Chechkin_Sokolov_Renewal,Evans_majumdar_JPhysA_review}. Here, we allowed the resetting rate to explicitly depend on the trajectory, making $T$ and $R$ statistically dependent, but showed that the same equations hold. The case of regular resetting is obtained trivially by omitting the state dependence from the resetting rate in equation (\ref{probRlet}). 

\subsection*{Estimating the MFPT from trajectories without resetting}

We now want to estimate how different state- and time-dependent resetting protocols affect the MFPT. Since analytical expressions are unavailable in almost all cases, one approach is to directly sample trajectories with resetting and estimate the MFPT. If one is interested in a single resetting protocol, that may be the most efficient approach. However, if one would like to explore a wide range of resetting protocols, e.g. for design and optimization purposes, this approach becomes impractical. Each resetting protocol would require resampling the trajectories from scratch which is very costly. We instead propose a much more efficient approach, based on equation (\ref{ISR MFPT TET}). Importantly, we show that sampling a single set of trajectories with no resetting is enough to estimate the MFPT for any state- and time-dependent resetting protocol through a reweighing procedure. 

We begin with a set of  $N$ trajectories $\boldsymbol{X}^i(t)$, where $i=1,2,\dots ,N$. We decompose each trajectory into $n_i$ time steps of length $\Delta t$. For $\Delta t$ that is sufficiently short, the probability of resetting the $i-$th trajectory at the $j-$th time step, given that no resetting occurred previously, is $p^i_j= r\big(\boldsymbol{X}^i(j\Delta t),j\Delta t\big)\Delta t$.
Then, we define the survival probability of trajectory $i$ up to time step $j$, i.e., the probability that no resetting occurred prior to time step $j$, as
\begin{equation}\label{Pi FPT}
    \Psi^i_j=\prod_{k=1}^{j-1}\left(1-p^i_k\right).
\end{equation}
Using the survival probability, we can estimate the success probability as
\begin{equation}\label{success probability}
    \begin{split}
        \Pr(T\leq R)&\approx\frac{1}{N}\sum_{i=1}^N \Psi^i_{n_i},
    \end{split}
\end{equation}
and the mean duration of a successful segment ending in first-passage as \begin{equation}\label{cMFPT}
    \begin{split}
        \langle T|T\leq R\rangle &\approx\frac{1}{N\Pr(T\leq R)}\sum_{i=1}^N \Psi^i_{n_i}n_i\Delta t.
    \end{split}
\end{equation}
Equation (\ref{success probability}) is simply the average probability of surviving the entire trajectory without resetting, and equation (\ref{cMFPT}) evaluates the average trajectory duration reweighed by its survival probability. Similarly, we can estimate the mean duration of a failed segment ending in resetting as
\begin{equation}\label{mean R}
    \langle R|R< T\rangle\approx\frac{1}{N(1-\Pr(T\leq R))}
    \sum_{i=1}^N\sum_{j=1}^{n_i-1}\Psi_{j}^i\,p_j^i\,j\Delta t,
\end{equation}
where $\Psi_{j}^i \, p_j^i$ is the probability that trajectory $i$ survives without resetting $j-1$ steps and is reset exactly at time step $j$.
Substituting equations (\ref{success probability}-\ref{mean R}) into equation (\ref{ISR MFPT TET}), we obtain an estimation for the MFPT whose accuracy is limited only by the sample size $N$ and the time step $\Delta t$.

In SI Fig. S1, we benchmark this procedure by comparing its result with the analytical MFPT for diffusion with asymmetric resetting, obtained in ref.\cite{Asymmetric_SR} We find excellent agreement between them, using sets of only $N=10^4$ trajectories without resetting from numerical simulations. The estimated MFPT values are up to $\sim 3\%$ off the analytic solution in almost all cases, for resetting rates in a range spanning over three orders of magnitude. Full simulation details are given in the SI.

 \section*{Search with environmental information} 
 
 We now employ our framework to study a biologically motivated example. Imagine a diffusive searcher in one dimension, with a diffusion coefficient $D$. The searcher starts at $L$, and the target is located at the origin. Without resetting, the MFPT for this search process diverges.~\cite{redner_book} However, Evans and Majumdar showed that using standard resetting, the MFPT becomes finite. \cite{Evans_majumdar_PRL}. 
As illustrated in Fig.~\ref{fig: intro}a, the searcher may further reduce the MFPT by incorporating information about the distance from the target, avoiding resetting too close to it. To model this environmental information, we employ a position-dependent resetting rate that increases up to a maximal value $r_0$ as a function of the distance from the target,
 \begin{equation}\label{eq: information senstivity}
     r(x)=r_0\left(1-\frac{1}{1+\left(x/b\right)^2}\right),
 \end{equation}
where $b$ is the typical length scale at which environmental information affects the searcher.

\begin{figure}[t]
     \centering
     \includegraphics[width=\linewidth]{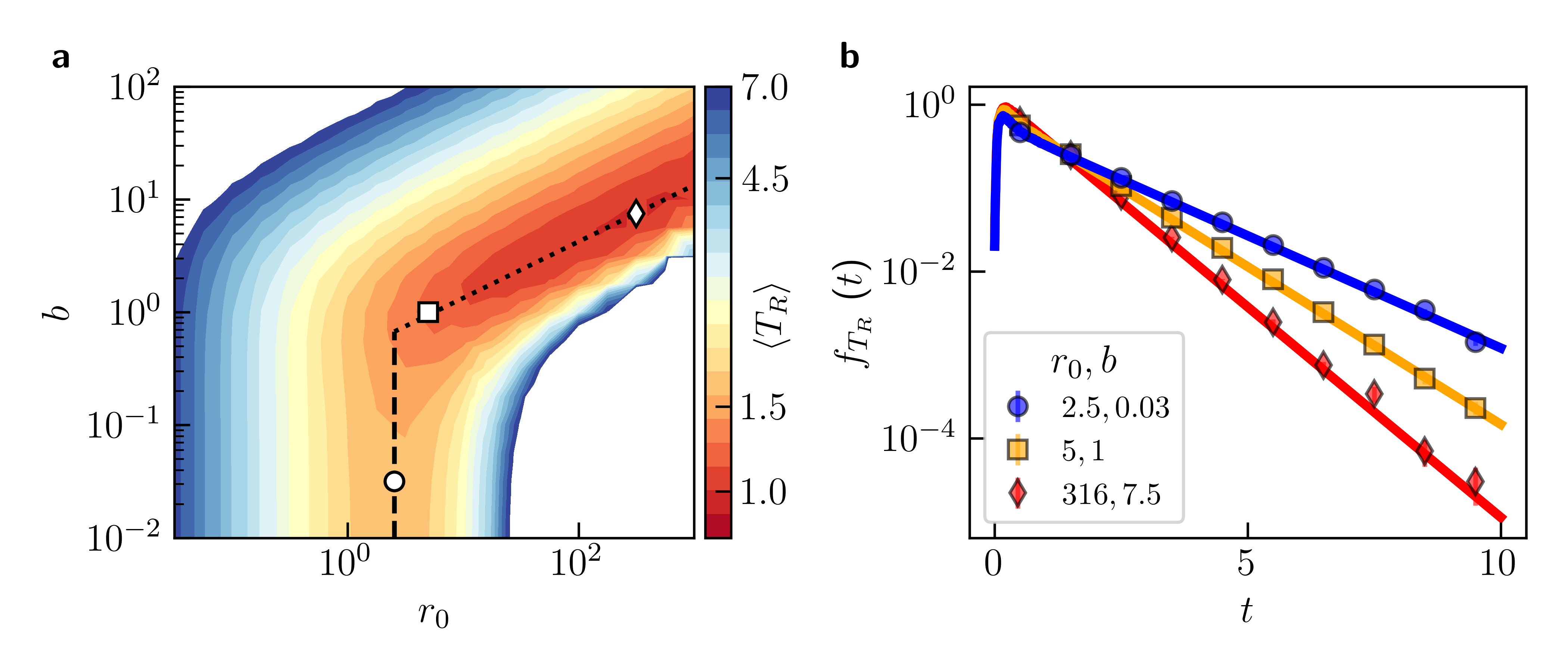}
 \caption{\textbf{a} The MFPT as a function of $r_0$ and $b$ for one-dimensional diffusion from $x=1$ to the origin under position-dependent resetting of the form given in equation (\ref{eq: information senstivity}). The dashed and dotted lines indicate the optimal rate for diffusion with a constant resetting rate, $r_0=2.5$, and the optimal rate for a parabolic resetting, $r_0=5.6b^2$, respectively. \textbf{b} FPT distributions for the three marked points in the phase space of panel \textbf{a}, estimated from simulations without resetting (solid lines), and brute-force simulations with resetting (markers). For simulations with resetting, we plot the mean $\pm 1/\sqrt{10}$ standard deviation of ten independent repetitions with $10^4$ trajectories each (errors are smaller than marker size in almost all cases).}
     \label{fig:bio}
 \end{figure}

 Fig. \ref{fig:bio}a presents the MFPT as a function of the control parameters $r_0$ and $b$.
We observe that there are two regimes. In the limit of $b \ll L$, there is no information effectively and the behavior is like free diffusion with standard resetting at a constant rate $r_0$. The minimum of the MFPT is obtained at the expected rate for free diffusion~\cite{Evans_majumdar_PRL}, indicated by the black dashed line. However, when $b \gg L$, equation (\ref{eq: information senstivity}) is approximated by  $r(x)\simeq r_0/b^2 \,x^2$, and resetting is governed by a single parameter $r_0/b^2$ that determines the curvature of the parabolic resetting rate. When the curvature is very small, the problem asymptotically converges to free diffusion, where the MFPT diverges. When the curvature is very large, the particle is trapped at the vicinity of $L$, and the MFPT again diverges. Therefore, we expect that there will be an optimal value of the curvature, but there is currently no analytical expression for it. Using our method, we find the optimal value that leads to the highest acceleration, $r_0/ b^2 \simeq 5.6$ (black dotted line), which results in a MFPT of $\simeq 1$.

We emphasize that all the results of Fig.~\ref{fig:bio}a were obtained using a single set of trajectories with no resetting and simply re-evaluating equations (\ref{Pi FPT}-\ref{mean R}) for every value of $r_0$ and $b$, leading to a different MFPT through equation (\ref{ISR MFPT TET}). 
Previously, obtaining the above results required performing an ensemble of brute-force simulations at every value of the parameters $r_0$ and $b$, to map out the MFPT phase space. This approach would be computationally prohibitive in many cases. Moreover, the strength of our approach becomes even clearer when considering other forms of $r(x)$ apart from equation (\ref{eq: information senstivity}). Tackling these using brute-force simulations, analytical methods, or experiments, would require complete reanalysis of the problem. On the other hand, using our method, we can use the \textit{same} initial ensemble of trajectories to generate the MFPT phase space for any resetting protocol, with minimal added cost. In the SI, we demonstrate this for several other sigmoidal-shaped resetting protocols.

By efficiently extracting the MFPT phase space for several functional forms of the resetting rate we identify an interesting trend. For all forms, there are two regimes. The behavior at $b \ll L$ mimics that of diffusion with a constant resetting rate, and the behavior at $b \gg L$ is determined entirely by the first-order term in the Taylor expansion of the resetting rate near the origin. All other particularities of the resetting rate only play a role at the crossover regime $b \approx L$.

Finally, we are not limited to the MFPT and can obtain the full FPT distribution for any value of $r_0$ and $b$ using the same single set of trajectories. Fig. \ref{fig:bio}b shows the FPT distributions obtained for the three values marked in Fig. \ref{fig:bio}a, as estimated using the set of trajectories without resetting (solid lines), and from direct simulations with resetting for comparison (markers). We find excellent agreement between the two for all cases. The derivation of how to estimate the FPT distribution from trajectories without resetting is given in the SI.
 
\section*{Estimating the Propagator and NESS} 

We next show how to evaluate the full propagator of the process under state- and time-dependent resetting from a single set of trajectories without resetting. The long time limit of the propagator provides the NESS achieved by the resetting protocol. 

We begin by defining the propagator, $G_j(\vec{x})$, as the probability to be at $\vec{x}$ at time step $j$. It has two contributions, the first from trajectories that are at $\vec{x}$ in time step $j$ without undergoing resetting before. The second contribution is from trajectories that underwent resetting at least once. Overall, the propagator can be written as,   
\begin{equation}\label{eq: renewal propagator}
    G_j(\Vec{x})=G_j^{\Psi}(\Vec{x})+\sum_{k=1}^{j-1}P_kG_{j-k}(\Vec{x}),
\end{equation}
where $G_j^{\Psi}(\Vec{x})$ is the probability of being at position $\Vec{x}$ at time step $j$ without undergoing resetting along the way. The second term considers scenarios in which trajectories were reset for the first time at time step $k$, and $j-k$ steps later are at $\vec{x}$. Here,  $P_k $ is the probability of undergoing resetting for the first time at time step $k$.

To solve equation (\ref{eq: renewal propagator}), we write it in matrix form and rearrange, to get
\begin{equation}\label{eq: vectorized propagator}
    \Vec{G}(\Vec{x})=\left(\boldsymbol{I}-\boldsymbol{P}\right)^{-1}\Vec{G}^{\Psi}(\Vec{x}),
\end{equation}
where the matrix elements of  $\boldsymbol{P}$ are given by $\boldsymbol{P}_{i,j}\equiv P_{i-j}$ if $i>j$ and zero otherwise. Thus, to estimate the propagator under state- and time-dependent resetting, we only need to estimate $\boldsymbol{P}$ and $\Vec{G}^{\Psi}(x)$ from trajectories without resetting. We do this as follows:  
\begin{equation}
    \boldsymbol{P}_{i,j}=\begin{cases}
        P_{i-j}=\langle \Psi_{i-j} p_{i-j} \rangle \approx \frac{1}{N}\sum_{n=1}^N \Psi_{i-j}^n\, p_{i-j}^n & \text{if $i>j$},\\
        0 &\text{otherwise,}
    \end{cases}
\end{equation}
and
\begin{equation}\label{eq: G Psi evaluation}
    G^{\Psi}_j(\Vec{x})\approx\frac{1}{N}\sum_{i=1}^N\Psi_j^i\delta(\Vec{x}-\boldsymbol{X}^i(j \Delta t)).
\end{equation}
The advantage of this approach is that the solution is obtained in real-time, however, for long times, it requires inverting very large matrices. 

Alternatively, the long-time behavior and the NESS can be obtained more efficiently by taking the Z-transform of equation (\ref{eq: renewal propagator}) and using the convolution theorem. This gives
\begin{equation}
    \hat{G}(\Vec{x},z)=\frac{\hat{G}^\Psi(\Vec{x},z)}{1-\hat{P}(z)},
    \label{PropZtransform}
\end{equation}
where $\hat{f}(z)=\sum_{j=0}^\infty f_jz^j$ is the Z-transform of the series $\{f_0,f_1,f_2,..\}$. While equation (\ref{PropZtransform}) is given in discrete time, an equivalent equation using the Laplace transform is valid for continuous time. We note that a special case of the continuous-time analog of equation (\ref{PropZtransform}), for an overdamped particle diffusing in a potential and space-dependent resetting rate, was given by Roldán and Gupta.~\cite{PI_formalism_for_SR} 
Our work extends this result to a general stochastic process. 

The final value theorem for Z-transforms states that $G_{NESS}(\Vec{x})=\lim_{z\to1^-}(1-z)\hat{G}(\Vec{x},z)$. By using it, we get  (see SI)
\begin{equation}\label{eq:NESSthroughZ}
    G_{NESS}(\Vec{x})=\frac{\hat{G}^\Psi(\Vec{x},1)}{\langle N_R\rangle},
\end{equation}
where $\langle N_R\rangle$ is the mean number of time steps between consecutive resetting events. Note that the steady state in equation (\ref{eq:NESSthroughZ}) is well defined whenever $\langle N_R \rangle$ is finite, regardless of whether or not the process without resetting has a steady-state. This is a generalization of a well-known results in the theory of standard resetting to state- and time-dependent resetting~\cite{Pal_2016, Nagar_2016, Evans_majumdar_JPhysA_review}. 

To estimate the NESS, we first sample a set of $N$ trajectories without resetting of length $M\Delta t$. We stress that $M$ should be large enough such that, had we used resetting, the probability of surviving $M$ steps without resetting would be negligible, i.e., $\Psi_M^i\ll 1\,\forall i$. Then, we use equations (\ref{eq: G Psi  evaluation}) and (\ref{eq:NESSthroughZ}), and the definition of the Z-transform, to obtain

\begin{equation}\label{eq: NESS}
    G_{NESS}(\Vec{x})\propto \hat{G}^\Psi(\Vec{x},1) = \sum_{j=1}^\infty G^{\Psi}_j(\Vec{x})\approx\frac{1}{N}\sum_{j=1}^M\sum_{i=1}^N\Psi_j^i\delta(\Vec{x}-\boldsymbol{X}^i(j \Delta t)).
\end{equation}
This estimation results in an unnormalized distribution, which can be normalized easily. The normalization factor will be an estimation for the mean time between consecutive resetting events, $\langle N_R \rangle$.
Equation (\ref{eq: NESS}) shows that the estimation of the NESS with resetting, from trajectories without resetting, is done by averaging the histogram of positions over time and trajectories, but reweighing each trajectory, at every time step, by its survival probability.

\section*{Prediction and design of non-equilibrium steady states} 

The above results can be used to predict and design NESS of spatially-dependent resetting protocols. We demonstrate this using two examples.

 It is well known that for free diffusion with a constant resetting rate, a Laplace distributed NESS emerges \cite{Evans_majumdar_PRL}. 
 An analytical solution for the NESS of diffusion with a parabolic resetting rate $r(x)=r_0x^2$ is also known~\cite{PI_formalism_for_SR}.
 Interestingly, in both cases, the tails of the NESS decay as $\sim e^{-|x|^\alpha}$, with $\alpha=1$ for the constant resetting rate, and $\alpha=2$ for the parabolic resetting rate. This raises a more general question: what is the asymptotics of the NESS for diffusion with a power-law resetting rate $r(x)=r_0|x|^\lambda$. While there are currently no known closed-form solutions for the NESS with $\lambda \neq\{0,2\}$, we can easily estimate the resulting NESS using the procedure described in the previous section. 
 
 \begin{figure}[t]
     \centering
     \includegraphics[width=\linewidth]{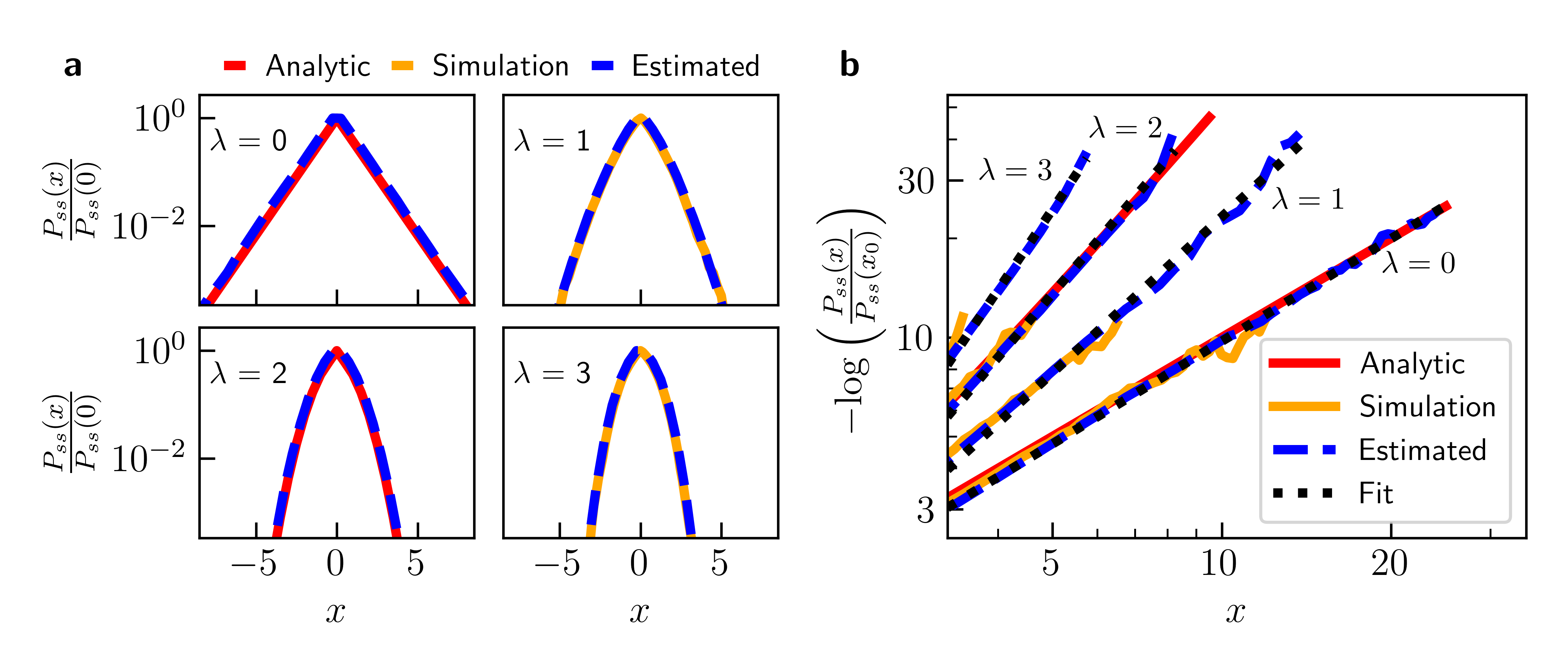}
 \caption{NESS for one-dimensional diffusion with a resetting rate $r(x) = r_0 x^\lambda$, for $r_0 = 1$ and $\lambda = \left\{0,1,2,3 \right\}$. Analytic solutions are given in red, results from brute-force simulations in yellow, and estimations based on trajectories with no resetting in blue. The black dotted lines in panel \textbf{b} give linear fits in the region $x > 3.25 \,$.}
     \label{fig:steadyState}
 \end{figure}

Fig. \ref{fig:steadyState}a presents the NESS for diffusion with a resetting rate $r(x)=r_0|x|^\lambda$, for $\lambda = \{0,1,2,3\}$. Red solid lines indicate the analytical solution when it is known. When analytical solutions are not known, we present results of brute-force Langevin dynamics simulations with stochastic resetting (yellow solid lines) instead. The blue dashed lines are obtained using the estimation method developed in the previous section, based on a single set of free diffusion trajectories without resetting. In all cases, we observe a good agreement between our method and the ground truth (theory or brute-force simulations). 

In Fig. \ref{fig:steadyState}b we show that the NESS asymptotics obeys $\sim e^{-|x|^\alpha}$ and estimate the values of $\alpha$ through linear regression. This procedure retrieves the correct values for $\lambda=\{0,2\}$, where we estimate $\alpha \simeq\{1.04,2.00\}$, respectively.   
For $\lambda=\{1,3\}$, we estimate  $\alpha\simeq\{1.57,2.53\}$, respectively. Following these results, we hypothesize that for a resetting rate $r(x)\sim |x|^\lambda$, the tails of the NESS decay as $\sim e^{-|x|^{\frac{\lambda}{2}+1}}$. While this hypothesis remains to be proven, it exemplifies the power of our approach, revealing new phenomena and inspiring future work.

Next, we demonstrate how to use our approach to engineer desirable NESS of interest. We do so by designing the arrow-shaped steady state, that is illustrated in Fig. \ref{fig: intro}b. We start with a system composed of a diffusing particle within a box with reflective boundaries at $x=\{-5,5\}$ and $y=\{-5,5\}$. Restart takes the particle to a uniformly distributed initial position within the box. The resulting NESS, for a position-independent resetting rate, is uniform. We seek to design a position-dependent resetting rate that will generate the desired arrow-shaped NESS, using only a single set of trajectories without resetting.

\begin{figure}[t!]
     \centering
     \includegraphics[width=\linewidth]{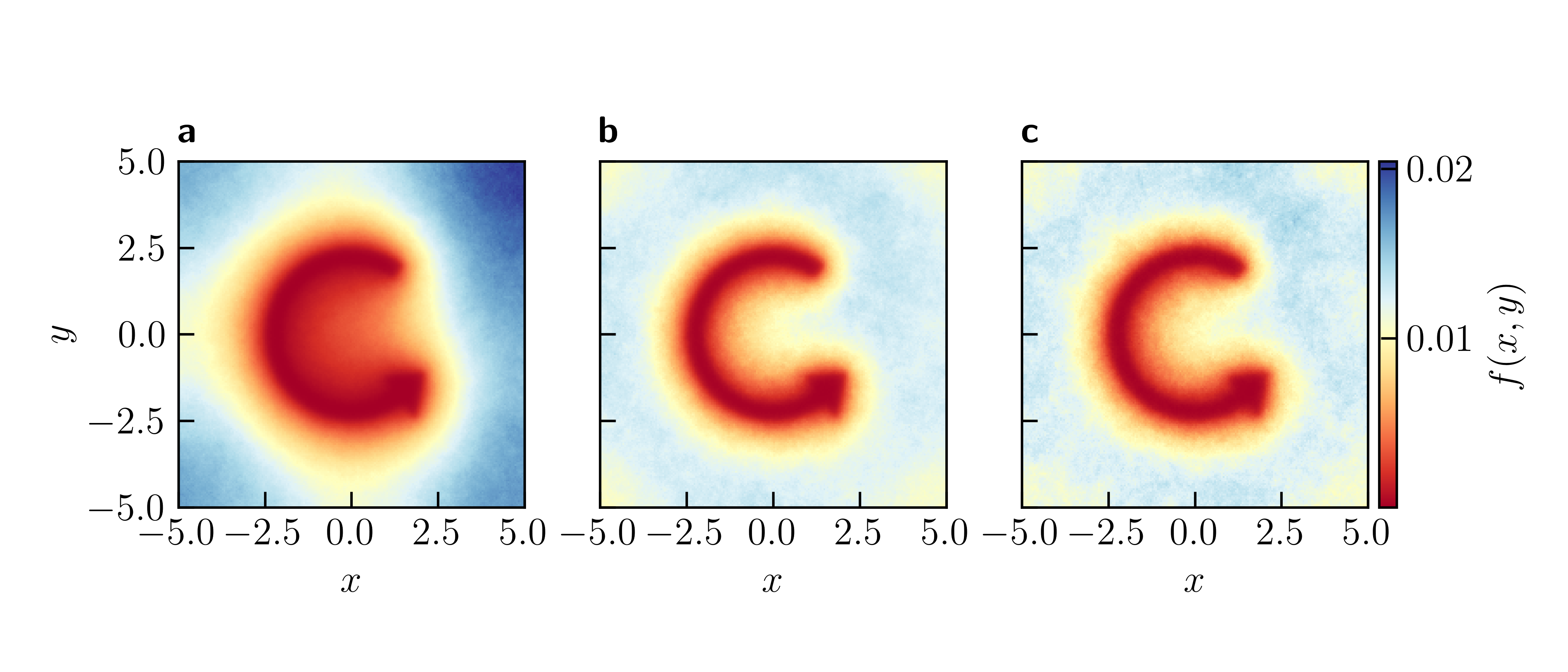}
 \caption{Two-dimensional arrow-shaped NESS. \textbf{a} predicted from simulations without resetting with a naive $r(x,y)$. \textbf{b} predicted from simulations without resetting with an improved $r(x,y)$. \textbf{c} obtained from brute-force simulations with the improved resetting protocol.}
     \label{fig:steadyStateDesign}
 \end{figure}

A naive approach would be to simply apply a constant resetting rate within the desired shape and zero resetting rate outside (see SI). However, using our approach to predict the resulting steady-state, we find that this naive guess leads to a very fuzzy distribution, in which the arrow is barely discernible (see Fig. \ref{fig:steadyStateDesign}a). This problem cannot be solved by increasing the resetting rate. The reason is that, for every reasonably complicated NESS, there will be areas that are very hard to reach without crossing regions with a high resetting rate. In our case, the area engulfed by the arrow can only be approached from the right, which significantly lowers its occupation probability, leading to the fuzzy arrow.

Previously, testing different resetting protocols, in a trial-and-error fashion, would have required running thousands of trajectories with resetting for every protocol until the desired steady-state would have been obtained. Instead, using our approach, we can design an improved state-dependent resetting protocol (see SI) that would lead to the desired shape (Fig. \ref{fig:steadyStateDesign}b), using the same set of trajectories without resetting that is already available.
Results from simulations with this resetting protocol are given in Fig. \ref{fig:steadyStateDesign}c for comparison, showing excellent agreement with our prediction.

\section*{Neural network optimization of the resetting protocol}
To conclude this paper, we show how to systematically optimize an adaptive resetting protocol for an arbitrary task based on our theory. We do so by representing the adaptive resetting probability $p\big(\boldsymbol{X},t\big)= r\big(\boldsymbol{X}, t)\Delta t$ as a neural network. Then, we define a loss function for training which can be any observable of the process with the adaptive resetting protocol given by the neural network. For example, this could be any function of the MFPT, FPT distribution, propagator, and NESS. The training data is the trajectories without resetting and the value of the loss function at every epoch is given by the predicted value of the desired observable under resetting using our theory. Optimization is carried out using standard techniques, such as stochastic gradient descent (see SI for the full details). We demonstrate the usefulness of our machine learning framework to optimize the adaptive resetting probability for minimizing the MFPT of conformational transitions of a protein in molecular dynamics simulations. 

Molecular dynamics simulations are a powerful tool, but their accessible timescales are limited to a few microseconds. Therefore, to simulate any physical phenomena in longer timescales, e.g., protein folding and crystal nucleation, one must use enhanced sampling methods~\cite{torrie_nonphysical_1977,kastner_umbrella_2011,sugita_replica-exchange_1999,faradjian_computing_2004,elber_milestoning_2020,barducci_metadynamics_2011,valsson_enhancing_2016,invernizzi_rethinking_2020}. Stochastic resetting recently emerged as a promising technique for that purpose, either as a standalone method~\cite{Ofir_JPCL,non_exponential_kinetics} or in combination with Metadynamics~\cite{Ofir_NatureCom,Church2024} -- a popular enhanced sampling tool~\cite{barducci_metadynamics_2011,valsson_enhancing_2016}. Resetting accelerated rare events in molecular dynamics and Metadynamics simulations by more than an order of magnitude and provided an accurate inference of the kinetics of the unperturbed process. However, in almost all cases, the resetting rate employed was state-independent. 
We recently showed that even a very limited functional form of adaptive resetting, designed ad hoc, substantially lowered the MFPT~\cite{Church2024}. 
We now show that representing the rate by a flexible neural network allows automatic optimization and yields new adaptive resetting protocols that result in higher speedups than previously possible. 

\begin{figure}[t!]
     \centering
     \includegraphics[width=\linewidth]{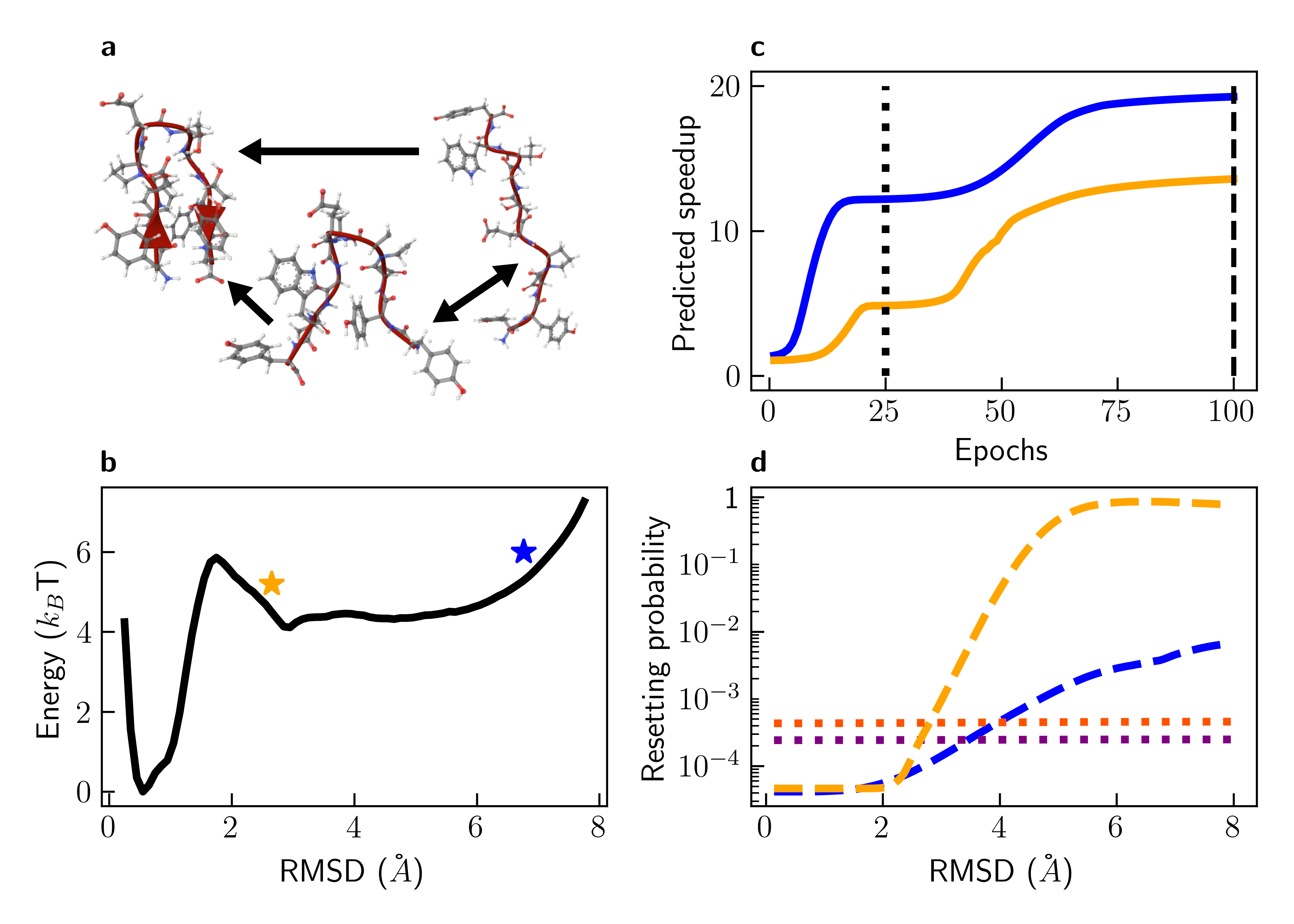}
 \caption{\textbf{a} Ball-and-stick representation of the folded (left), misfolded (center), and unfolded (right) states of chignolin in explicit water solvent (5889 atoms), including a cartoon representation of the backbone in crimson. The white, gray, blue, and red spheres represent hydrogen, carbon, nitrogen, and oxygen atoms, respectively.
 \textbf{b} The free energy along the RMSD. The RMSD values at the unfolded and misfolded states are indicated with blue and yellow stars, respectively. \textbf{c} The predicted MFPT under a neural network-based resetting protocol as a function of the number of epochs. \textbf{d} Neural network-based resetting probability as a function of the RMSD. Results in panels \textbf{c}, \textbf{d} are given for simulations initiated at the unfolded state (blue) or the misfolded state (yellow).}
     \label{fig:chignolin}
 \end{figure}

 We will demonstrate our approach by finding the resetting strategy that leads to the highest speedup (lowest MFPT) for the folding of the chignolin mini-protein in explicit water, which consists of 5889 atoms (full simulation details are given in the SI).
The system has three metastable states: an unfolded state, a misfolded state, and the folded, native state. Representative configurations of the states are given in Fig. \ref{fig:chignolin}a. We identify the states using the C-alpha root-mean-square deviation (RMSD) from the folded configuration. The free energy along this degree of freedom is plotted in Fig. \ref{fig:chignolin}b. The blue and yellow stars mark the RMSD values of the unfolded and misfolded configurations of Fig. \ref{fig:chignolin}a, respectively. We observe that there is an energy barrier the system has to cross in order to get to the native state (the deep well around RMSD $< 1.5 \, \AA$). There is no substantial barrier between the misfolded and unfolded states, such that multiple misfolding-unfolding events often occur before a successful folding. We consider two first-passage processes leading to the folded state from two initial configurations: the unfolded and misfolded states.

 To optimize the resetting strategy, we built a network with a simple architecture: It receives three inputs (the RMSD, the radius of gyration, and the end-to-end distance of the protein) and has two inner layers with 10 nodes each. The output is the probability to reset given the three inputs, restricted to be between zero and one using a sigmoid activation function in the final layer. The network is trained by going over data of trajectories with no resetting and estimating the MFPT under the resetting probability represented by the network using Eqs.
 (\ref{ISR MFPT TET}-\ref{mean R}). 
 The estimated MFPT serves as a loss function that is minimized during training. An example script to train the model is given on GitHub.

Fig. \ref{fig:chignolin}c shows the training curves of the model as a function of the number of epochs when starting from the unfolded (blue) or misfolded (yellow) states. The plotted value is the estimated speedup, which is defined as the ratio of the MFPT values without and with resetting, respectively. For both curves, there is a plateau around 25 epochs (highlighted with a dotted vertical line), where the algorithm reached a local minimum. The dotted lines in Fig. \ref{fig:chignolin}d show the resetting protocol represented by the network at this epoch as a function of the RMSD, averaged over the other two degrees of freedom. Purple and orange colors represent simulations initiated at the unfolded and misfolded states, respectively. In both cases, the networks suggest state-independent resetting probabilities. Remarkably, these are the resetting probabilities we identified as providing the highest acceleration for state-independent resetting in Refs~\cite{Ofir_NatureCom,Church2024}.

In later stages of training, the networks find better, state-dependent resetting protocols, reaching a second plateau around the 100th epoch, highlighted with a dashed vertical line in Fig. \ref{fig:chignolin}c. The corresponding adaptive resetting probabilities are plotted with dashed lines in Fig. \ref{fig:chignolin}d. For simulations initiated at the misfolded state, the predicted speedup is very close to the one obtained in Ref.~\cite{Church2024} using an \textit{ad hoc}, naive functional form. The resetting probability represented by the network has a similar functional form, starting with a small probability that increases sharply to nearly $\sim1$ for RMSD values larger than $\sim 6 \, \AA$~\cite{Church2024}. Most importantly, for simulations initiated at the unfolded configuration, the naive functional form did not lead to any speedup, but our machine learning optimization framework finds a protocol that grows gradually with the RMSD and leads to higher speedups. 

\section*{Conclusions and Outlook}
To conclude, we presented a general formulation of state- and time-dependent stochastic resetting protocols, which we call adaptive resetting.
Our formulation generalizes key results from the theory of resetting for the MFPT, FPT distribution, propagator, and NESS. We presented a numerical scheme to predict all these fundamental properties using a \textit{single} set of trajectories without resetting. 
We focused on resetting protocols that depend only on the current state and age of the system. However, our results apply without change to resetting protocols that depend on previous states and even the entire history of the trajectories. 
We demonstrated the power of our approach through several examples ranging from investigating informed search strategies to predicting and designing NESS. 
 
We also developed a machine learning framework to optimize adaptive resetting strategies for an arbitrary task for the first time. We showed that our scheme can discover non-trivial resetting protocols for accelerating molecular simulations (lowering the MFPT) beyond the previous state of the art of state-independent resetting. In the future, the same approach can be used to optimize other observables, such as the entire FPT distribution or the NESS.

With the broader capabilities that adaptive resetting offers, we anticipate it will advance research on core problems in enhanced sampling. In particular, we anticipate that adaptive resetting will be harnessed for the accurate evaluation of equilibrium rates and free-energies from non-equilibrium protocols~\cite{Tiwary2013, Salvalaglio2014, Tiwary2015, Dellago2014, elber_milestoning_2020, kuznets-speck_inferring_2023, blumer_short-time_2024}. Similarly, we expect it will soon be brought into play for the estimation of rates and free energies in NESS~\cite{warmflash_umbrella_2007, RosaRaices2024, Heller2024} and for investigating first-passage dynamics out of equilibrium~\cite{militaru2021}. Overall, we believe that adaptive resetting will emerge as a powerful and versatile tool, opening new avenues for research and applications in non-equilibrium statistical mechanics, molecular simulations, and beyond.

\bibliography{references}
\end{document}


\section{Comparison between the Numerical and Analytical result of the MFPT for Asymmetric Resetting}\label{SI: asymmetric resetting}

To benchmark the procedure for the MFPT estimation, we compared our result to the one obtained analytically for the problem of asymmetric resetting presented in ref~\cite{Asymmetric_SR}. In this problem, a particle diffuses in 1D with a diffusion coefficient $D$ from $L>0$ to a target located at the origin. For $x>L$, the particle is subjected to stochastic resetting with rate $r_0$, which reset it to $L$. In our notation, we write the position-dependent resetting rate as $r(x)=r_0\Theta(x-L)$, where $\Theta(y)$ is the Heaviside step function.

\begin{figure}[t]
     \centering
     \includegraphics[width=0.5\linewidth]{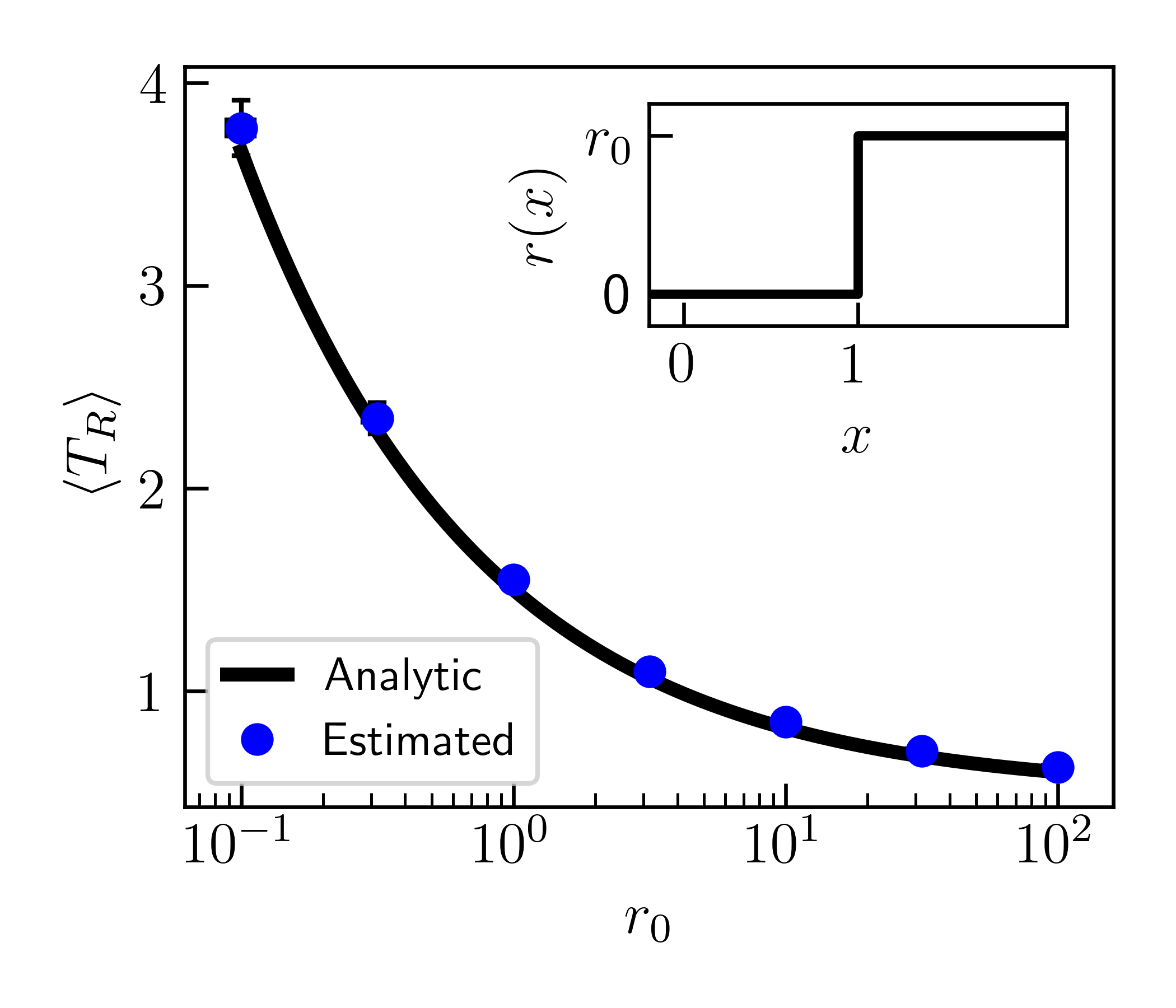}
 \caption{The MFPT as a function of $r_0$ for one-dimensional diffusion of a particle initiated at $x=1$ with absorbing wall at $x=0$, under an asymmetric resetting protocol (inset). The analytical solution is given in black. The blue circles show the average prediction obtained from 100 bootstrapping batches of $10^4$ trajectories without resetting. The boxes and whiskers show the IQR and extreme values within 1.5 IQR of the first and third quartiles, respectively (they are smaller than marker size in almost all cases). Time is measured in units of $L^2/D$.}
     \label{fig:firstPassage}
 \end{figure}

The MFPT under this resetting protocol is $\langle T_R\rangle=L^2/2D+L/\sqrt{r_0D}$ \cite{Asymmetric_SR}. We compared this analytical result to the one obtained using the proposed method, for $D=1$ and $L=1$. We simulated $10^5$ trajectories without resetting and obtained 100 bootstrapping batches, each containing $10^4$ randomly sampled trajectories. Each batch provided an estimation of the MFPT. Fig. \ref{fig:firstPassage} displays a boxes and whiskers plot of the bootstrapping results, with the boxes marking the range between the first and third quartiles (interquartile range, IQR), and the whiskers marking extreme values within 1.5 IQR below and above these quartiles. The mean of the estimations is plotted with blue circles. In most rates, the boxes are smaller than the circles' size.

\section{Numerical simulations}

Here, we provide the details of all numerical simulations presented in the manuscript and the following section, except for the molecular simulations of chignolin.
In most cases, we simulated overdamped Langevin dynamics of a single particle in one dimension, through the discretization
\begin{equation}\label{eq:discretization}
x(t) = x(t-1) + \sqrt{2D \Delta t} \eta,
\end{equation}
with diffusion coefficient $D=1$, time step $\Delta t=0.001$, and $\eta$ being Gaussian noise with a mean of 0 and standard deviation of 1. For Fig. 5c of the main text, we simulated two-dimensional dynamics, with each coordinate obeying equation (\ref{eq:discretization}).

Fig. 3 of the main text uses data from $10^4$ trajectories initiated at $x=1$ and halted when reaching $x \le 0$. Figs. 4-5 of the manuscript use data from $10^4$ trajectories initiated at $x=0$, including $10^5$ time steps each. Fig. 4b uses $\lvert x_j^i \rvert$, with $x_j^i$ being the value of $x$ at time step $j$ of trajectory $i$. 

For Fig. 5, we generated $10^5$ two-dimensional trajectories using the $10^4$ one-dimensional trajectories, through the following algorithm:
\begin{itemize}
\item Sample a random trajectory $i$ to serve as the $x$-coordinate and a random trajectory $k \ne i$ to serve as the $y$-coordinate.
\item Add random constants $C_x, C_y \in [-5,5]$ to all $x_j^i$ and $x_j^k$, respectively.
\item To implement reflecting boundary conditions, scan the trajectories: if at time step $j'$ you find $x_{j'}^i>5$ or $x_{j'}^i<-5$, add $10 -2 x_j^i$ or $-10 -2 x_j^i$ to all $x_{j\ge j'}^i$, respectively. Do the same for trajectory $k$.
\item Save $\left(x^i,x^k\right)$ as a new two-dimensional trajectory.
\end{itemize}

Finally, Fig. 5c uses a two-dimensional trajectory of $10^8$ time steps with the position-dependent resetting protocol of equation (\ref{eq:resettingProtocol}).

\section{Search with environmental information}

 \begin{figure}[t!]
     \centering
     \includegraphics[width=\linewidth]{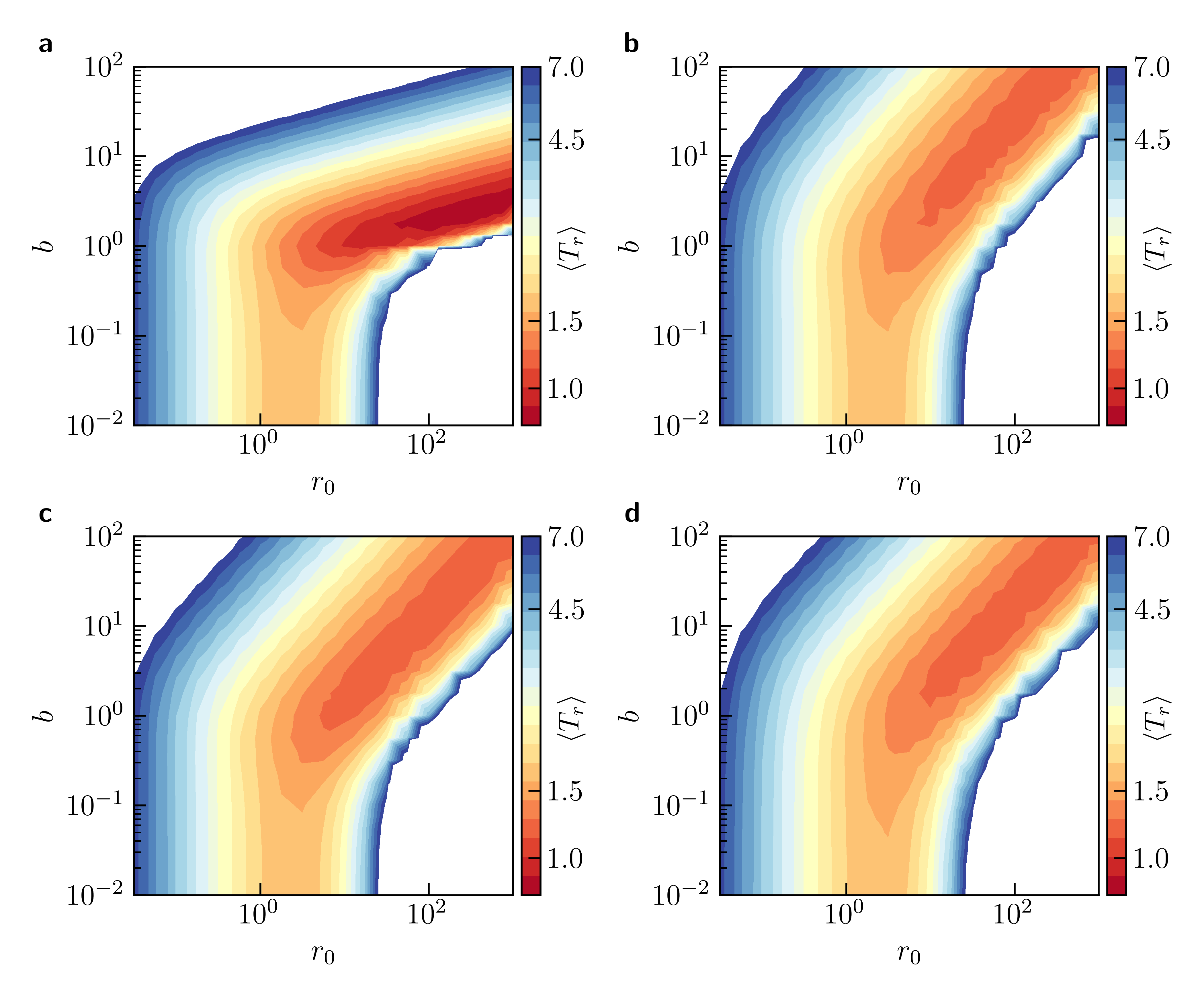}
 \caption{The MFPT as a function of $r_0$ and $b$ for one-dimensional diffusion from $x=1$ to the origin under position-dependent resetting of the form given in (\textbf{a}) equation (\ref{eq:info2}), (\textbf{b}) equation (\ref{eq:info3}), (\textbf{c}) equation (\ref{eq:info4}), and (\textbf{d}) equation (\ref{eq:info5}).}
     \label{fig:differentForms}
 \end{figure}

Fig. \ref{fig:differentForms} shows the MFPT of the one-dimensional search as a function of parameters $r_0$ and $b$ for different functional forms:
 \begin{equation}\label{eq:info2}
     r(x)=r_0\left(1-\frac{1}{1+\left(x/b\right)^4}\right),
 \end{equation}
\begin{equation}\label{eq:info3}
     r(x)=r_0\left(1-\exp\left({-\frac{x}{b}}\right)\right),
 \end{equation}
\begin{equation}\label{eq:info4}
     r(x)=r_0\left(\frac{2}{1+\exp(-x/b)}-1\right),
 \end{equation}
\begin{equation}\label{eq:info5}
     r(x)=\frac{2r_0}{\pi}\arctan\left(\frac{x}{b}\right).
 \end{equation}

\section{Estimating the FPT Distribution}
We now show how to estimate the FPT distribution with resetting. We define $F_j$ as the probability for first-passage to occur at the $j-$th time step. This probability must obey the following renewal equation

\begin{equation}\label{eq: renewal FPT distribution}
    F_j=f^{\Psi}_j+\sum_{k=1}^{j-1}P_kF_{j-k},
\end{equation}
where $f^{\Psi}_j$ is the probability of first-passage at time step $j$ without undergoing resetting along the way, and $P_k$ is the probability for a reset to occur for the first time at time step $k$. The second term in equation (\ref{eq: renewal FPT distribution}) accounts for trajectories that were reset at the $k$-th time step, and then managed to finish in $j-k$ time steps, resulting in a first-passage after $j$ time steps overall.

Equation (\ref{eq: renewal FPT distribution}) can be solved either by writing it in a matrix form, or in Z-space. The advantage of the first method is that the solution is obtained in real-time, but for long times it requires inverting large matrices. The second method provides a solution in Z-time, from which the long-time behavior of the FPT distribution can be easily obtained.

Writing equation (\ref{eq: renewal FPT distribution}) in matrix form results in
\begin{equation}
    \Vec{F}=\left(\boldsymbol{I}-\boldsymbol{P}\right)^{-1}\Vec{f^\Psi},
\end{equation}
where the matrix elements of $\boldsymbol{P}$ are given by $\boldsymbol{P}_{i,j}\equiv P_{i-j}$ if $i>j$ and zero otherwise.

For the solution in Z-space, we take the Z-transform of both sides of equation (\ref{eq: renewal FPT distribution}), where the Z-transform is defined as $\hat{f}(z)=\sum_{j=0}^\infty f_jz^j$. We then use the convolution theorem for Z-transforms to obtain

\begin{equation}
    \hat{F}(z)=\frac{\hat{f}_{\Psi}(z)}{1-\hat{P}(z)}.
\end{equation}
The long time behaviour of $F_j$ can be obtained by looking at $\hat{F}(z)$ in the vicinity of $z=1$.

Using a set of $i=1,2,...N$ trajectories of the process without resetting, each ending in a first passage after $n_i$ time steps, it is possible to estimate both $f_j^\Psi$ and $P_k$. This can be done in the following way,
\begin{equation}
    \begin{split}
        f_j^\Psi&\approx \frac{1}{N}\sum_{i=1}^N \Psi_{n_i}^i\delta_{n_i,j},\\
        P_k&\approx \frac{1}{N}\sum_{i=1}^N p_k^i\,\Psi_k^i,
    \end{split}
\end{equation}
where $\delta_{i,j}$ is the Kronecker delta.

In the case where the MFPT diverges, waiting for all trajectories to show first-passage is not feasible. In that case, one should gather trajectories until first-passage or until some time-step $M\gg 1$. The estimation of the FPT distribution will be reliable if for all unfinished trajectories $\Psi^i_M\ll1$. The error in the FPT distribution can be estimated using bootstrapping.

\section{Derivation of Equation 14 in the Main Text}

The propagator of the process with resetting in Z-time is given by equation (13) in the main text. The NESS is defined as $G_{NESS}(\Vec{x})=\lim_{j\to\infty}G_j(\Vec{x})$ if this limit exists, and the integral of the resulting function over $\Vec{x}$ is normalized. This limit can be obtained using the final-value theorem for Z-transforms
\begin{equation}
    G_{NESS}(\Vec{x})=\lim_{z\to1^-}(1-z)\hat{G}(\Vec{x},z)=\lim_{z\to1^-}\frac{(1-z)\hat{G}^\Psi(\Vec{x},z)}{1-\hat{P}(z)}=\hat{G}^\Psi(\Vec{x},1)\lim_{z\to1^-}\frac{1-z}{1-\hat{P}(z)}.
\end{equation}
To compute this limit we will use the moments expansion of the Z-transform around $z=1$, which reads $\hat{f}(z)=1+(z-1)\langle f\rangle+o(z-1)$. Where $\langle f\rangle$ is the mean of the probability mass function, generated by $\hat{f}(z)$. Using it on $\hat{P}(z)$ to evaluate the limit gives
\begin{equation}
    G_{NESS}(\Vec{x})=\hat{G}^\Psi(\Vec{x},1)\lim_{z\to1^-}\frac{1-z}{1-1+(1-z)\langle N_R\rangle}=\frac{\hat{G}^\Psi(\Vec{x},1)}{
    \langle N_R\rangle},
\end{equation}
where $\langle N_R\rangle$ is the mean number of time steps between consecutive resetting events.

\section{Arrow-shaped non-equilibrium steady state}

\begin{figure}[t!]
     \centering
     \includegraphics[width=\linewidth]{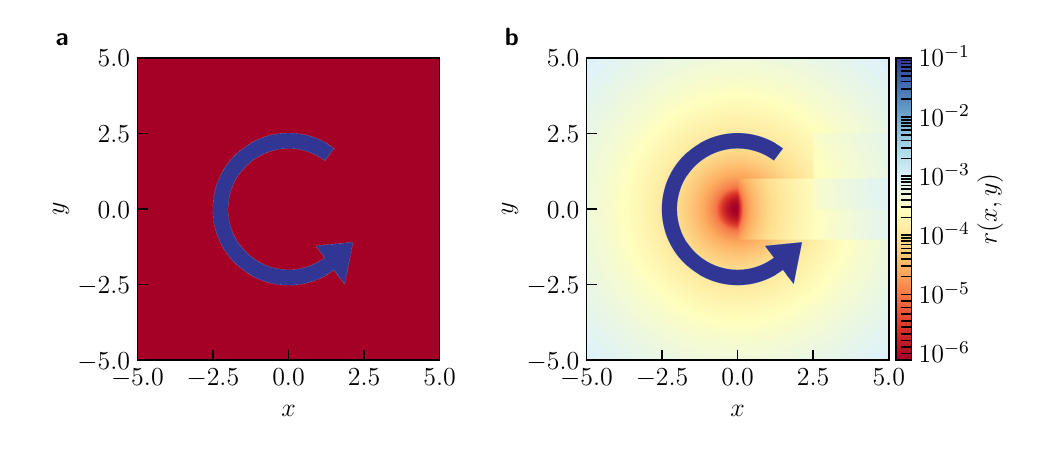}
 \caption{Resetting rates given by (\textbf{a}) equation (\ref{eq:resettingProtocol}) and (\textbf{b}) equation (\ref{eq:resettingProtocol2}).}
     \label{fig:resettingProtocol}
 \end{figure}

Here, we provide the position-dependent resetting rates that produce the NESS presented in Fig. 5 of the main text. Equation (\ref{eq:resettingProtocol}) is used for panel \textbf{a} of the figure, while equation (\ref{eq:resettingProtocol2}) is used for panels \textbf{b} and \textbf{c}. Equations (\ref{eq:resettingProtocol}) and (\ref{eq:resettingProtocol2}) are plotted in panels \textbf{a} and \textbf{b} of Fig.~\ref{fig:resettingProtocol}, respectively.

\begin{equation}\label{eq:resettingProtocol}
r(x,y)=\begin{cases}
			100, & \begin{split}
   \text{if} 
\left[\left(\sqrt{x^2+y^2}>2\right) \wedge \left(\sqrt{x^2+y^2}<2.5\right) \wedge \left(x<1.5\right)\right] \cup \\
\left[\left(-\frac{4}{3}x-y < 0\right) \wedge \left(0.1x-y> 1.3\right) \wedge \left(5x-y < 11.7\right)\right]\end{split}\\
            0, & \text{otherwise}.
		 \end{cases}
\end{equation}

\begin{equation}\label{eq:resettingProtocol2}
r(x,y)=\begin{cases}
			100, & \begin{split}
   \text{if} 
\left[\left(\sqrt{x^2+y^2}>2\right) \wedge \left(\sqrt{x^2+y^2}<2.5\right) \wedge \left(x<1.5\right)\right] \cup \\
\left[\left(-\frac{4}{3}x-y < 0\right) \wedge \left(0.1x-y> 1.3\right) \wedge \left(5x-y < 11.7\right)\right]\end{split}\\

            \begin{split}0.02\sqrt{x^2+y^2} + \\0.1\end{split}, & \text{if} \left(x>2.5\right) \wedge
            \left(1<y<2.5\right) \\

            \begin{split}0.02\sqrt{x^2+y^2} + \\0.05x+0.1\end{split}, & \text{if} \left(x>2.5\right) \wedge
            \left(0<y<1\right) \\

            \begin{split}0.02\sqrt{x^2+y^2} + \\0.05x\end{split}, & \text{if} \left[\left(0<x<2.5\right) \wedge
            \left(0<y<1\right)\right]
            \cup \left[\left(x>0\right) \wedge
            \left(-1<y<0\right)\right]\\
            
            0.02\sqrt{x^2+y^2}, & \text{otherwise}.
		 \end{cases}
\end{equation}

\section{Neural networks architecture}

Here, we provide the details of the architecture used to minimize the MFPT of conformational transitions
in molecular simulations.
The neural networks were designed and trained using Pytorch. Both were composed of three fully connected layers. The first layer received three inputs by the user and produced 10 outputs. The second received 10 inputs and produced 10 outputs, and the third received the 10 outputs and produced a single output. The inner layers were separated by ReLU activation functions, and the output layer was followed by a sigmoid activation function. 

We present results after 25 or 100 training epochs. The weights and biases were initialized such that the output is $\sim 0$ for all inputs, and the resetting protocol represented by the networks is never to reset. We minimized the loss function through stochastic gradient descent with no momentum. A thousand trajectories with no resetting were used as training data, and all data was used at each epoch. Learning rates of $0.00001$ and $0.000015$ were used for data from simulations initiated at the misfolded and unfolded states, respectively. An example of the training code is available on the GitHub repository.

\section{Simulations of solvated chignolin}

Here, we provide details of the simulations of chignolin we used as training data for the neural networks described in the manuscript. We used trajectories of Metadynamics~\cite{barducci_metadynamics_2011,valsson_enhancing_2016} simulations, introducing bias along a collective variable based on the C-alpha root-mean-square deviation from a folded configuration, at a pace of once every 100 timesteps. The simulations were carried out in GROMACS 2019.6~\cite{abraham_gromacs_2015}, with Metadynamics implemented through PLUMED 2.7.1.~\cite{bonomi_plumed_2009,tribello_plumed_2014,Bonomi2019}. The simulations setup was the same as in previous works~\cite{Ofir_NatureCom,Church2024}, and additional information is detailed there.

\bibliography{references}